\documentclass[11pt,twoside]{article}
\usepackage{asp2010}

\resetcounters

\begin{document}

\title{Accelerating the Rate of Astronomical Discovery with GPU-Powered Clusters}
\author{Christopher~J.\ Fluke
\affil{Centre for Astrophysics \& Supercomputing, Swinburne University 
of Technology, Hawthorn, Victoria, Australia}}

\begin{abstract}          
In recent years, the Graphics Processing Unit (GPU) has emerged
as a low-cost alternative for high performance computing, enabling
impressive speed-ups for a range of scientific computing applications.
Early adopters in astronomy are already benefiting in adapting their
codes to take advantage of the GPU's massively parallel processing paradigm.
I give an introduction to, and overview of, the use of GPUs in
astronomy to date, highlighting the adoption and application trends
from the first $\sim\!\!100$ GPU-related publications in astronomy. 
I discuss the opportunities and challenges of utilising GPU computing clusters,
such as the new Australian GPU supercomputer, gSTAR, 
for accelerating the rate of astronomical discovery.
\end{abstract}

\section{Introduction}
For at least four decades from the 1960s, advances in traditional
computation on single-core CPUs has been driven by increases
in transistor density and clock rate. This is seen through the well-established
Moore's Law \citep{Moore} biennial doubling in the number of transistors 
per integrated circuit, and a corresponding increase in processing performance. 
In principle, it was possible to
implement a code once, and achieve faster (approximately double)
computation simply by purchasing new hardware, at lower cost, every two years.
In practice, new generations of CPUs also provided additional
benefits (such as increased system memory, improved caching, etc.),
resulting in on-going algorithmic improvements and software updates.

In the early 2000s, CPU clock-rates began to plateau -- mainly  due to manufacturing 
constraints, such as difficulties in keeping ever-faster CPUs sufficiently cool to work 
without melting.  Further processing improvements, and the 
continuation of Moore's Law growth, were achieved 
by moving to multi-core solutions.  Indeed, the likely future of CPUs is that they
will become increasingly multi-core: codes or algorithms that can be
expressed in parallel form will derive the most benefit from these
new architectures.  A preview of this highly multi-core future is 
available now in the guise of the many-core graphics processing unit (GPU). 
Leveraging advances in hardware that were designed to enhance and improve 
graphical performance in support of the many-billion dollar international 
computer gaming industry,  GPUs have rapidly become credible 
alternatives for low-cost, massively parallel scientific computation.
Astronomers have been quick to adopt GPUs as a powerful new component 
of their computational arsenal.  

Following early successes at speeding-up codes on single GPU systems and 
small-scale GPU clusters, a growing number of research 
institutions are now making major investments in significant high-performance 
computing (HPC) clusters deriving a substantial fraction of their (theoretical) 
peak processing performance from GPUs.  
At the dawn of this exciting new era of GPU-powered HPC clusters, 
what do astronomers need to know about GPUs in order to take advantage of 
new computational opportunities?  
What does a potential 10x or 100x processing speed-up mean in terms 
of accelerating the rate of astronomical discovery?  What lessons 
can we learn, and what trends can we identify, 
from the early adopters of GPUs in astronomy?  And now that we have 
$O(100)$ Tflop/s GPU-clusters at our disposal,\footnote{We use the notation 
Gflop/s = $10^9$ floating point operations per second and Tflop/s = $10^{12}$ 
flop/s.} just what are we going to do with them?

\section{GPUs for Scientific Computation}
In essence, the GPU acts as a computational co-processor to the CPU,
a mode of operation not unfamiliar to computer programmers (and owners) of
the 1980s who could opt to use a maths co-processor or floating point unit (FPU)
to accelerate mathematical operations. While modern GPUs offer 
a much wider range of programmable capability than the earlier FPUs, 
they are not able to completely replace the CPU -- nor are they likely to.  
In general terms, GPUs achieve their performance at the hardware level 
by trading off the large-memory caches and sophisticated control logic 
of CPUs (accommodating software solutions for activities 
as diverse as opening a file from a local disk, serving a web-page in a 
browser, and numerical processing for an astrophysical simulation) 
for circuit-area devoted to fast floating point computations.  

As the potential for using the highly-parallel GPU architecture for scientific
computation became apparent \citep{2003cs.......10002V}, the notion
of general purpose computation on graphics processing units (GPGPU) began
to gain momentum.  Early attempts to utilise the increased 
computational performance of GPUs required programming in shader 
languages [e.g. NVIDIA's C for Graphics, Cg, was used by 
\citet{2004SPIE.5572..262R} and \citet*{2007NewA...12..641P}]. 
For graphics, the hardware processing pipeline is optimised to calculate 
red-green-blue (RGB) colours and alpha (A) channel transparency for pixels, 
vertices and polygons, achieved through the use of customised 
software fragment shader functions.  Implementation of a scientific algorithm was 
only possible if it could be  recast as a shader, 
often requiring storing data in structures that shared  
the ``four floating point numbers'' structure of RGBA. 

The advent of the Compute Unified Device Architecture 
(CUDA\footnote{\url{http://www.nvidia.com/cuda}}) application programming 
interface (API) from NVIDIA and the open-standard 
alternative OpenCL,\footnote{\url{http://www.khronos.org/opencl}} developed by 
the Khronos Group, have dramatically changed the usability of the GPU
for general computation.\footnote{For more details on GPU programming, see,
e.g. \citet{KirkHwu} or \citet{SandersKandrot}.}
Indeed, certain GPU products from vendors,
such as the Tesla series from NVIDIA, are sold with scientific
computing in mind:  architecturally equivalent to consumer graphics 
hardware, they lack the capacity to output graphics to a display device,
but with increased memory spaces and provision for error correcting memory, which are
not required by the home computer gamer.
Moreover, while early generations of GPUs only supported 32-bit 
(single precision) floating bit computation, the higher-end solutions 
now also provide 64-bit (double precision) support at comparable
processing speeds.  

\section{Early Adopters and Emerging Trends}
One of the first astronomical problems adapted to GPU was acceleration of the
$N$-body force problem, through computation of the $O(N^2)$ pair-wise forces
between particles.  Early GPU implementations were reported by 
\citet{Nyland04}, using Cg and OpenGL on an NVIDIA GPU, 
while \citet{Elsen06} and \citet{2007arXiv0706.3060E} used 
BrookGPU \citep{Buck04} on an ATI X1900XTX card.  
Both groups found that the high arithmetic intensity of the force calculation 
was ideally suited to the GPU's architecture, and simple code optimisations 
could give speed-ups of more than $20\times$ compared to existing 
CPU implementations. Moreover, 
they achieved computational performance comparable to the more expensive, 
custom GRAPE-6A hardware.

\citet{2004SPIE.5572..262R} examined a real-time problem - recovery of  
the wave-front phase from a Shack-Hartmann sensor.  Reporting 
on an implementation of the iterative Hudgin algorithm, they found 
a 10x speed-up for the centroid part of the calculation, but only a 2x 
speed-up overall compared to a CPU-only implementation.
They demonstrated that peak performance on a GPU does require a 
sufficiently large problem - the CPU out-performs the GPU when there 
are insufficient processing tasks to keep the GPU pipeline busy.

\citet{2004ExA....17..287S} described a Common-Off-The-Shelf (COTS) 
correlator platform constructed from GPUs, with an eye on future, 
low-cost solutions scalable to the Square Kilometre Array (SKA).  
They achieved $\sim\!\!5\times$ better performance (measured as
complex multiplications/second) for a $16 \times$ larger problem 
using an NVIDIA GeForce 6800 Ultra GPU, compared with a 2.8 GHz CPU.
The price/Gflop and power usage/Gflop of the GPU were both about 3x better 
than for CPU.  

To examine some of the emerging trends in the adoption of GPUs 
in astronomy since these early projects, we perform 
simple bibliometrics using the SAO/NASA Astrophysics Data System 
Abstract Service.\footnote{\url{http://adsabs.harvard.edu}}   An 
abstract-only search on various combinations of the terms: {\em GPU(s)}, 
{\em  graphics processing unit(s)}, {\em CUDA}, and {\em OpenCL} resulted 
in 94 distinct abstracts from 2004-2011 (as of 1 October 2011). There 
were no relevant abstracts in 2005.  An attempt was made to remove 
duplicate items (e.g. papers that appear separately as an arXiv version 
and a final published version).

There are, of course, limitations with such an approach. We fail to 
identify those publications that used GPUs, but did not declare this in 
the abstract \citep*[e.g.][]{2010arXiv1012.2901F}, and not all publications 
on astronomy-related GPU applications appear in ADS 
[e.g. \citet{10.1109/SC.2010.1}; \citet{spurzem2010}]. While
additional details on the API and specific hardware used could have been obtained 
from each of the publications, our intention was to obtain a quick snapshot
of the current state of GPU development and the extent of early adoption 
in astronomy. We can answer questions such as ``how are GPUs being used in
astronomy?'' (Figure 1) and ``where are the results being published?'' (Figure 2). 

Analysis of the abstracts reveals almost 50 unique computational problems 
in 30 broad application areas, ranging from adaptive optics 
and algorithm analysis, data mining and 
digital signal processing, plasma and protoplanetary disk simulation, 
to tree-codes and two-point correlation functions.  The vast majority 
of abstracts present GPU-based codes or methods (82/94).  In this 
context a ``method'' relates 
to a demonstration that a particular problem is suited to a GPU, and 
is often accompanied by a quoted speed-up (relative to a single-core CPU,
or, in a small number cases, a multi-core CPU implementation) or a peak 
processing performance  
[e.g.  \citet{2007astro.ph..3100H}; \citet*{2008NewA...13..103B}; 
\citet*{2010arXiv1005.5384G}].  Of the remaining abstracts, 9 were clearly 
identifiable as presenting new scientific results based on the 
use of an existing GPU code  [e.g. \citet{2010ApJ...724..244A}; 
\citet*{2010MNRAS.402..371B}; \citet*{2011MNRAS.tmp.1539G}],  
and three dealt more generally 
with the ``philosophy'' of adopting GPUs for scientific computing in astronomy
[e.g. \citet*{2010MNRAS.408.1936B}].

\begin{figure}[!ht]
\plotone{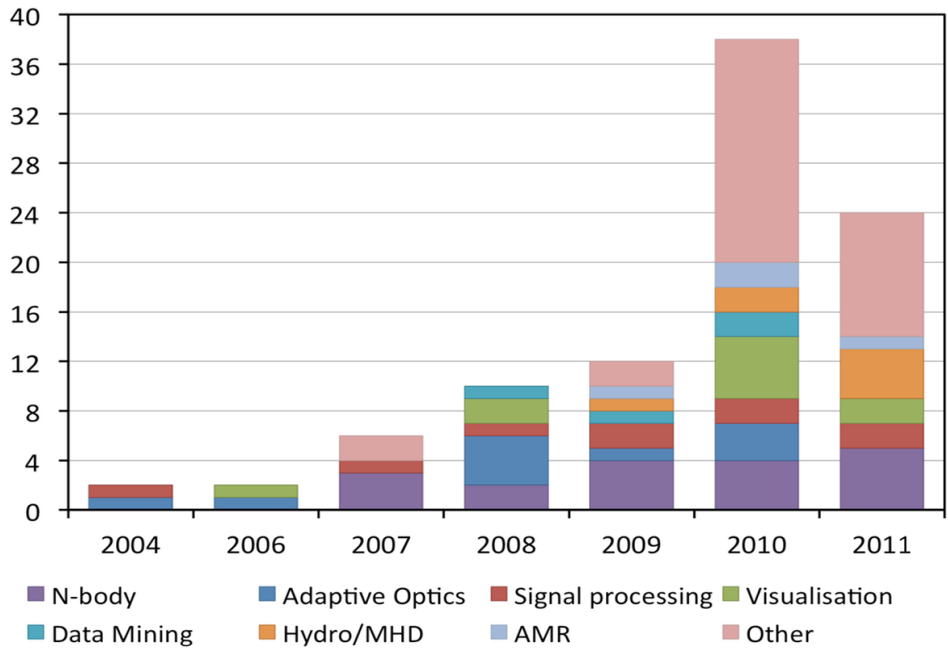}
\caption{{\em How are GPUs being used in astronomy?}
An ADS abstract-only search (1 October 2011) on various 
combinations of the terms: {\em GPU(s)}, {\em  graphics processing unit(s)}, 
{\em CUDA}, and {\em OpenCL} resulted in 94 publications (there
were no relevant abstracts in 2005).  The ``other''
category combines any application area with 3 or fewer abstracts.  
The year 2010 represents the commencement of wider adoption of GPUs by the 
astronomical community. } 
\label{I05-fig-1}
\end{figure}

As Figure 1 shows, the year 2010 marked the transition from early 
exploration of the capabilities and suitability of GPUs to 
a restricted number of problems, to one of widespread adoption 
across a broad range of application areas in astronomy 
(62 abstracts across 26 application areas since 2010 -- the ``other''
category combines any application area with 3 or fewer abstracts).  
We anticipate that 
this trend will continue for at least the next few years, as 
the application market is far from being saturated.   
Amongst the early applications, Fourier transforms and pair-wise N-body 
forces  were obvious, ``low-hanging fruit'', 
with straight-forward parallelism. Recent works are tackling 
more complex algorithms, such as general relativistic 
magnetohydrodynamics \citep{2011arXiv1102.5202Z}.

N-body simulations (and related methods) stand out as being the most 
popular target for both methods and scientific result abstracts (18/94).
The emphasis on scientific computing is clear, with only ten abstracts 
discussing visualisation/data analysis uses of GPUs -- a number comparable 
with GPU-enabled signal processing for radio astronomy (9/94), 
adaptive optics (10/94) and hydrodynamics/magnetohydrodynamics (7/94).

\begin{figure}[!ht]
\plotone{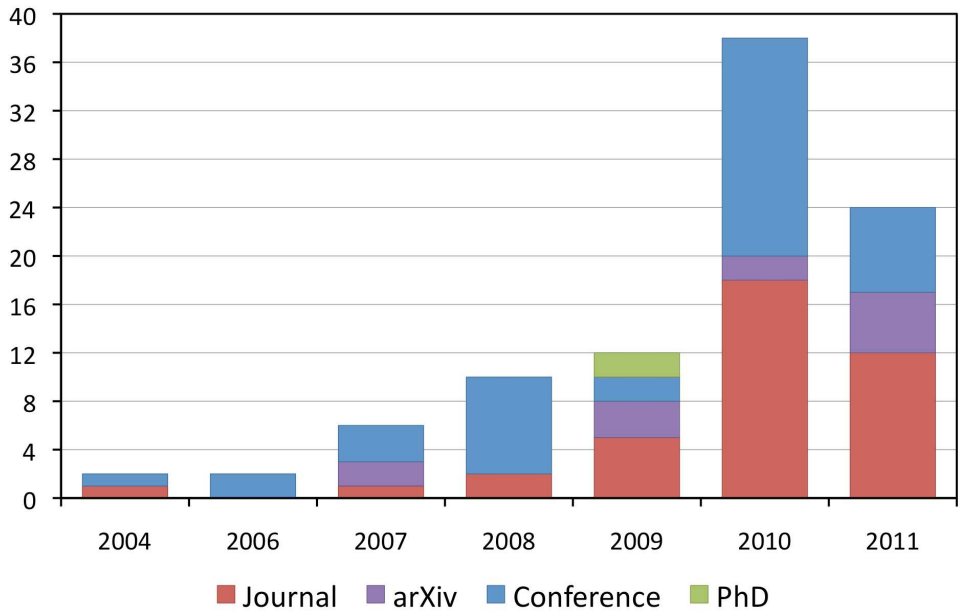}
\caption{{\em Where is GPU-related work being published?} Journals include New Astronomy, Monthly Notices of the Royal Astronomical Society, Astrophysical Journal, Astronomy \& Astrophysics, Publications of the Astronomical Society of Australia; Conferences include SPIE and ADASS; the arXiv 
category includes publications that are not clearly identifiable with 
one of the other categories.} \label{I05-fig-2}
\end{figure}

It is interesting to see where GPU-related work is being 
published -- see Figure 2.  We identify four categories of publication outlet:
journals, arXiv preprints, conference papers and PhD theses.
The arXiv category includes papers 
that have appeared on the arXiv, but are not clearly identifiable with 
one of the other categories (as of 1 October 2011).\footnote{e.g. papers 
that are identified as being in-press or accepted to a named journal 
are counted in the journal category, however, they are still counted towards
the year in which they first appeared on the arXiv.}
The main contributors to the conferences category are SPIE (11/94) -- mostly
presentations on adaptive optics -- and ADASS (6/94). The two main 
astronomy journals 
publishing refereed GPU-papers are New Astronomy (13/94) and 
Monthly Notices of the Royal Astronomical Society (7/94). 
The important message from this is that journals are prepared to
publish GPU methods papers, so get out there and turn your conference
papers into more complete publications with details of your algorithms
(and kernels), so that others can benefit from your experiences.

Based on abstracts only, other trends were considered. The declared use 
of particular programming APIs: Cg (2; none since 2007), 
CUDA (26; since 2008), and OpenCL (7; since 2010).  
17 abstracts noted the use of a specific NVIDIA card, with 
Tesla/Fermi cards increasing in prominence (NVIDIA S1070,  C1060, 
and C2050 cards are identified in six abstracts since 2010).  Only 
two abstracts named ATI cards (\citealt*{2010arXiv1004.1680P};
\citealt{2007arXiv0706.3060E}). At 
present, NVIDIA and CUDA do seem to dominate the scientific computing 
market share, which may be due in part to the more recent
appearance of OpenCL as a general-purpose programming API suitable
for ATI cards. 

Reported speed-ups, relative to CPU-implementations, ranged from 7x  
[computation of the Fast-Fourier Transform for adaptive optics in 
\citet{2006SPIE.6272E..35R}] to 600x [solving Kepler's equations in 
\citet{2009NewA...14..406F}], with several projects highlighting 
one-to-two order of magnitude improvements in performance 
[radio astronomy signal processing -- \citet*{2008ExA....22..129H}; 
magnetohydrodynamics -- \citet{2009arXiv0908.4362W}; cosmological 
lattice computations -- \citet{2010CoPhC.181..906S}].

We should treat speed-ups with some caution:  achieving performance at 
the highest end (100x) may be an indication that a less efficient, 
existing CPU-implementation was being compared with a
highly-optimised GPU solution.  Moreover, a single precision speed-up
is often more impressive than for double (or quadruple -- see 
\citealt{2010arXiv1006.0663G}) precision, so consideration must be 
made to accuracy over performance.
Devoting additional time to optimising existing CPU-codes is possibly 
not time well spent, particularly if a ``simple'' GPU code remains faster [see 
examples in \citet{2011PASA...28...15F}], however, an 
investigation of the potential for parallelisation of an existing 
single-core CPU code can lead to simple speed-ups through the use 
of libraries such as OpenMP\footnote{\url{http://openmp.org/}} on 
multi-core CPU architectures.  On the other hand, speed-ups reported 
even 1-2 years ago against single-core CPUs can comfortably be increased 
by a factor of a few: while GPU processing rates continue to grow, 
single-core CPU rates have stalled.

The current record holder for a ``workstation'' GPU (i.e. not a cluster, 
but still allowing for multiple GPUs within one device) is 1.28 Tflop/s 
on a Tesla S1070 for computing direct ray-tracing for microlensing 
\citep{2010NewA...15...16T}.   As with speed-ups, flop-counts should be 
treated with caution as there can be a mismatch between operations 
and clock-cycles when any mathematical operator beyond addition, 
subtraction or multiplication is used.

\section{GPU-Powered Clusters}
In terms of theoretical processing power, a single GPU can achieve the same
processing performance as a modest CPU cluster -- provided the problem can
fit in the memory of a single GPU --  and a small cluster of GPUs can 
outperform a $O(100)$-node CPU-based cluster, with a vast reduction in the 
amount of inter-node network connectivity required, and at a fraction of 
the hardware and operating cost.  In this context, a GPU cluster is 
really a hybrid CPU-GPU system, as GPUs cannot manage important tasks 
like reading data from disks or supporting networks.   

A growing number of astronomical institutions are now investing in major 
GPU-powered HPC clusters, with two of the first being the 
kolob compute cluster\footnote{\url{http://kolob.ziti.uni-heidelberg.de/}} 
at the University of Heidelberg, and the Silk Road 
facility\footnote{\url{http://silkroad.bao.ac.cn/}} operated by the National 
Astronomical Observatories of China \citep[see][]{spurzem2010}. 

In 2010, the Australian astronomy community was invited to present
expressions of interest to Astronomy Australia Limited (AAL) 
for new research infrastructure that could be funded by AAL through 
the Australian Federal Government's Education Investment Fund.  
One of the nine successful projects was gSTAR: the GPU Supercomputer for 
Theoretical Astrophysics Research -- a national high-performance 
computing facility for astronomers.  Installation of the gSTAR facility 
at Swinburne University of Technology commenced in early September 2011. 

Phase 1 of gSTAR, with a theoretical peak of $\sim\!\!130$ Tflop/s 
(single precision), comprises 51 dual-socket compute nodes each with
2 GPUS (NVIDIA C2070; 6 GB RAM), with an additional 3 high-density GPU 
compute nodes containing 7 GPUs (M2090: 6GB RAM).  In excess
of 1 Petabyte of usable disk space, supported by the Lustre file system, 
is available, and compute nodes are connected via QDR InfiniBand.  
Phase 2 of gSTAR, scheduled for 2012,  will see the addition of further 
GPUs.
Early science on gSTAR is expected to include: high-resolution N-body 
simulations of star clusters, including incorporation of improved 
physics; a cosmological microlensing parameter survey;
and extensions to the interactive, real-time visualisation framework
for terascale datasets \citep*{2011NewA...16..100H}. 

As of June 2011, 3 of the top 5 facilities (the National Supercomputing 
Centers in Tianjin and Shenzhen, 
and the GSIC Center, Tokyo Institute of Technology) on the TOP500
Supercomputing Sites\footnote{\url{http://www.top500.org/}} 
achieve their benchmark status, in part, through the use of GPUs.  These numbers
are likely to rise in the next TOP500 list.
Consulting the Green500 List,\footnote{\url{http://www.green500.org} (June 2011)}
which ranks HPC facilities based on their energy efficiency, 
4 of the top 10 use GPUs (ATI Radeon for Nagasaki University and 
Universitaet Frankfurt; NVIDA for GSIC Center, Tokyo Institute of Technology 
and CINECA/SCA - SuperComputing Solution). Moreover, 14 of the Top 20 Green sites
are accelerator based, using either GPUs or IBM Cell-based processors -- 
further evidence that GPUs can provide impressive energy efficiency. 

Astronomical use of GPU clusters to date has included adaptive-mesh-refinement
calculations \citep*[e.g.][]{2010ApJS..186..457S}; wavefront correction
for adaptive optics \citep{2009SPIE.7439E..11B};
spherical harmonic transforms for Cosmological Microwave
Background computations \citep{2011arXiv1106.0159S}; further 
progress on high-resolution, $N$-body simulations \citep{spurzem2010};
and real-time, interactive
volume rendering of terascale datasets \citep{2011NewA...16..100H}.

An innovative approach to gaining very high peak performance
was achieved by \citet{10.1109/SC.2010.1} through the use of low-end, 
commodity graphics cards (576 $\times$ NVIDIA GT200 cards).
A very impressive sustained performance of 190 Tflop/s for 
a 3 billion-particle, hierarchical $N$-body simulation, on a HPC 
cluster costing just over \$400,000 dollars, resulted 
in an honourable mention for the Gordon Bell Prize at Supercomputing 2010.  
This raises an interesting
comparison between high-end science cards and commodity GPUs - the main
difference is double precision performance, availability of error correcting
memory and total memory available.  If these are not concerns, then 
a commodity card may be sufficient.

\section{Accelerating the Rate of Discovery}
There are many reasons why astronomers might be excited about the
prospects of adopting GPUs. The most obvious benefits \citep[see also ][]{KirkHwu}
are the ability to:
\begin{itemize}
\item {\em Run an individual problem faster.} Computing a problem in a few 
minutes compared to a few days, or a few days compared to a few months; 
allows real-time solutions to computationally demanding tasks, such as 
detection of transient radio events [\citet{2011MNRAS.tmp.1518M}; 
\citet{Barsdell2011}].
\item {\em Run more problems in the same total wall time.}
Permits extensive exploration of parameter space  
[e.g. black hole inspirals -- \citet{2010CQGra..27c2001H}; 
solving Kepler's equations -- \citet{2009NewA...14..406F};
Lyman-$\alpha$ forest simulations -- \citet{2011MNRAS.tmp.1539G}].
This promises to be one of the most important new uses of GPU clusters, 
enabling greater understanding of the effects of initial conditions, and
allowing statistical investigations rather than promoting over-analysis of a 
single simulation result. 
\item {\em Solve a bigger problem size in the same wall time as a 
smaller problem size on a CPU-system.} This permits working 
at higher/improved resolution or provide greater capacity 
to explore evolution of systems over more time-steps; handle terascale,
and ultimately petascale, image and spectral data cube processing 
\citep{2010arXiv1012.2901F}, visualisation \citep{2011NewA...16..100H},
and data mining \citep{2010AAS...21523005P}.
However, if the problem cannot fit within the memory of a single GPU, 
a great deal of communication may be required between nodes of a GPU cluster. 
If the bottleneck moves from computation to data transfer, then gains
delivered by a processing speed-up may be lost until such time there is
a corresponding speed-up in bandwidth (between nodes), an increase in memory
bus size (between host and GPU), and a decrease in latency in the interconnect.
\item {\em Solve a more complex computational problem in the same wall time as a 
simpler problem on a CPU-system. }
E.g. use a more accurate solution method, which may exhibit better
stability, etc.; enable the inclusion of additional physical properties
(e.g. magnetic fields); opportunities to utilise/implement algorithms  
with improved accuracy rather than an increase in resolution or problem size.
\item {\em Provide much lower price/performance compared to an equivalent CPU-based cluster. } Provides potential for more astronomers to access Tflop/s high 
performance computing on the desktop, rather than needing to apply/compete 
for time on
national or institution-level HPC supercomputing facilities for all 
comp\-u\-tation\-ally-demanding processing.
\end{itemize}

The move from traditional 
CPU systems to GPU-accelerated computation is not without challenges.  
Identifying,  implementing, and optimising {\em relevant} algorithms 
for the highly-parallel GPU architecture can require a greater understanding
of computer science fundamentals than many science professional possess:
traditional sequential programming skills are arguably easier for
``astronomer-programmers'' to learn than parallel programming techniques.
More importantly, code that has been developed specifically for single-core CPU
will not run on a GPU without substantial modification.
In the short term, additional personnel time is required in 
order to develop GPU codes. 

In general, the best results on GPUs are seen for
computations that exhibit a large amount of data parallelism (i.e. the
same computation performed on many different data values) and
high arithmetic intensity (i.e. a high ratio of floating point calculations
to memory accesses). Learning to use a GPU effectively means gaining 
an understanding of a new range of programming tricks, including reducing
branching conditions (if/then statements), making judicious use of
over-computing (e.g. using zero-mass particles in the pairwise $N$-body 
force calculation) to keep GPU threads busy, and giving more thought 
to memory access patterns.

By placing the emphasis on the ``total time to science'' 
\citep{2011PASA...28...15F}, rather than time spent developing code 
for GPUs, some of this additional coding work should be made up by the 
typically $10\times$ (or greater) processing speed-ups.  As 
a growing number of GPU-programming and scripting libraries become available
(e.g.  
PyCuda\footnote{\url{http://mathema.tician.de/software/pycuda}} and
Thrust\footnote{\url{http://code.google.com/p/thrust/}}), with a
goal of improving developer productivity, the short-term need for new
code development may be reduced.   Interactive data languages such 
as IDL\footnote{\url{http://www.ittvis.com/}} can also achieve acceleration
through bindings to GPU libraries like 
GPULib,\footnote{\url{http://www.txcorp.com/products/GPULib/}} bringing the 
potential for GPU-acceleration to non-C-programming astronomers.

\section{Concluding Remarks}
At the dawn of the petascale data era, astronomers will be faced with new
challenges in data processing and computation. GPU-powered HPC clusters 
offer a low-cost opportunity to explore new, scalable, massively-parallel 
algorithms.  The processing speed-ups available with GPUs, for the 
right types of problems, are helping pave the way to new science, 
through higher-resolution simulations, improved physical modelling, 
and much greater exploration of parameter spaces.  Ultimately, the 
goal of adopting any new hardware solution in astronomy should be to 
help improve and enhance our understanding of the Universe. 
The future of computing for astronomy is here - and it is massively parallel.

\acknowledgements CJF has benefited greatly from GPU-related discussions 
with David Barnes, Ben Barsdell, and Amr Hassan.  This research has 
made use of NASA's Astrophysics 
Data System Bibliographic Services.

\bibliography{I05}

\begin{thebibliography}{}
\expandafter\ifx\csname natexlab\endcsname\relax\def\natexlab#1{#1}\fi
\expandafter\ifx\csname url\endcsname\relax
  \def\url#1{\texttt{#1}}\fi
\expandafter\ifx\csname urlprefix\endcsname\relax\def\urlprefix{URL }\fi
\providecommand{\eprint}[2][]{\url{#2}}

\bibitem[{{Aubert} \& {Teyssier}(2010)}]{2010ApJ...724..244A}
{Aubert}, D., \& {Teyssier}, R. 2010, \apj, 724, 244

\bibitem[{{Banerjee} et~al.(2010){Banerjee}, {Baumgardt}, \&
  {Kroupa}}]{2010MNRAS.402..371B}
{Banerjee}, S., {Baumgardt}, H., \& {Kroupa}, P. 2010, \mnras, 402, 371

\bibitem[{{Barsdell} et~al.(2011){Barsdell}, {Bailes}, {Barnes}, \&
  {Fluke}}]{Barsdell2011}
{Barsdell}, B.~R., {Bailes}, M., {Barnes}, D.~G., \& {Fluke}, C.~J. 2011,
  MNRAS, refereed

\bibitem[{{Barsdell} et~al.(2010){Barsdell}, {Barnes}, \&
  {Fluke}}]{2010MNRAS.408.1936B}
{Barsdell}, B.~R., {Barnes}, D.~G., \& {Fluke}, C.~J. 2010, \mnras, 408, 1936

\bibitem[{{Belleman} et~al.(2008){Belleman}, {B{\'e}dorf}, \& {Portegies
  Zwart}}]{2008NewA...13..103B}
{Belleman}, R.~G., {B{\'e}dorf}, J., \& {Portegies Zwart}, S.~F. 2008, NewA,
  13, 103

\bibitem[{{Bouchez} et~al.(2009)}]{2009SPIE.7439E..11B}
{Bouchez}, A.~H., et~al. 2009, in Society of Photo-Optical Instrumentation
  Engineers (SPIE) Conference Series, vol. 7439 of Society of Photo-Optical
  Instrumentation Engineers (SPIE) Conference Series

\bibitem[{Buck et~al.(2004)Buck, Foley, Horn, Sugerman, Fatahalian, Houston, \&
  Hanrahan}]{Buck04}
Buck, I., Foley, T., Horn, D., Sugerman, J., Fatahalian, K., Houston, M., \&
  Hanrahan, P. 2004, ACM Transactions on Graphics, 23, 777

\bibitem[{Elsen et~al.(2006)Elsen, Houston, Vishal, Darve, Hanrahan, \&
  Pande}]{Elsen06}
Elsen, E., Houston, M., Vishal, V., Darve, E., Hanrahan, P., \& Pande, V. 2006,
  in Proceedings of the 2006 ACM/IEEE conference on Supercomputing (New York,
  NY, USA: ACM), SC '06

\bibitem[{{Elsen} et~al.(2007){Elsen}, {Vishal}, {Houston}, {Pande},
  {Hanrahan}, \& {Darve}}]{2007arXiv0706.3060E}
{Elsen}, E., {Vishal}, V., {Houston}, M., {Pande}, V., {Hanrahan}, P., \&
  {Darve}, E. 2007, ArXiv e-prints. \eprint{0706.3060}

\bibitem[{{Fluke} et~al.(2011){Fluke}, {Barnes}, {Barsdell}, \&
  {Hassan}}]{2011PASA...28...15F}
{Fluke}, C.~J., {Barnes}, D.~G., {Barsdell}, B.~R., \& {Hassan}, A.~H. 2011,
  PASA, 28, 15

\bibitem[{{Fluke} et~al.(2010){Fluke}, {Barnes}, \&
  {Hassan}}]{2010arXiv1012.2901F}
{Fluke}, C.~J., {Barnes}, D.~G., \& {Hassan}, A.~H. 2010, ArXiv e-prints.
  \eprint{1012.2901}

\bibitem[{{Ford}(2009)}]{2009NewA...14..406F}
{Ford}, E.~B. 2009, NewA, 14, 406

\bibitem[{{Gaburov} et~al.(2010){Gaburov}, {B{\'e}dorf}, \& {Portegies
  Zwart}}]{2010arXiv1005.5384G}
{Gaburov}, E., {B{\'e}dorf}, J., \& {Portegies Zwart}, S. 2010, ArXiv e-prints.
  \eprint{1005.5384}

\bibitem[{{Ginjupalli} \& {Khanna}(2010)}]{2010arXiv1006.0663G}
{Ginjupalli}, R., \& {Khanna}, G. 2010, ArXiv e-prints. \eprint{1006.0663}

\bibitem[{{Greig} et~al.(2011){Greig}, {Bolton}, \&
  {Wyithe}}]{2011MNRAS.tmp.1539G}
{Greig}, B., {Bolton}, J.~S., \& {Wyithe}, J.~S.~B. 2011, \mnras, in press

\bibitem[{{Hamada} \& {Iitaka}(2007)}]{2007astro.ph..3100H}
{Hamada}, T., \& {Iitaka}, T. 2007, ArXiv e-prints. \eprint{astro-ph/0703100}

\bibitem[{Hamada \& Nitadori(2010)}]{10.1109/SC.2010.1}
Hamada, T., \& Nitadori, K. 2010, in Proceedings of the 2010 ACM/IEEE
  International Conference for High Performance Computing, Networking, Storage
  and Analysis (Washington, DC, USA: IEEE Computer Society), SC '10, 1

\bibitem[{{Harris} et~al.(2008){Harris}, {Haines}, \&
  {Staveley-Smith}}]{2008ExA....22..129H}
{Harris}, C., {Haines}, K., \& {Staveley-Smith}, L. 2008, ExA, 22, 129

\bibitem[{{Hassan} et~al.(2011){Hassan}, {Fluke}, \&
  {Barnes}}]{2011NewA...16..100H}
{Hassan}, A.~H., {Fluke}, C.~J., \& {Barnes}, D.~G. 2011, NewA, 16, 100

\bibitem[{{Herrmann} et~al.(2010){Herrmann}, {Silberholz}, {Bellone},
  {Guerberoff}, \& {Tiglio}}]{2010CQGra..27c2001H}
{Herrmann}, F., {Silberholz}, J., {Bellone}, M., {Guerberoff}, G., \& {Tiglio},
  M. 2010, CQG, 27, 032001

\bibitem[{{Kirk} \& {Hwu}(2010)}]{KirkHwu}
{Kirk}, D.~B., \& {Hwu}, W.-M.~W. 2010, {Programming Massively Parallel
  Processors: A Hands-on Approach} (Burlington, MA: Elsevier Inc)

\bibitem[{{Magro} et~al.(2011){Magro}, {Karastergiou}, {Salvini}, {Mort},
  {Dulwich}, \& {Zarb Adami}}]{2011MNRAS.tmp.1518M}
{Magro}, A., {Karastergiou}, A., {Salvini}, S., {Mort}, B., {Dulwich}, F., \&
  {Zarb Adami}, K. 2011, \mnras, in press

\bibitem[{{Moore}(1965)}]{Moore}
{Moore}, G.~E. 1965, Electronics, 38, 4

\bibitem[{{Nyland} et~al.(2004){Nyland}, {Harris}, \& {Prins}}]{Nyland04}
{Nyland}, L., {Harris}, M., \& {Prins}, J. 2004, in GP$^2$, The ACM Workshop on
  General Purpose Computing on Graphics Hardware, poster presentation

\bibitem[{{Pang} et~al.(2010){Pang}, {Pen}, \& {Perrone}}]{2010arXiv1004.1680P}
{Pang}, B., {Pen}, U.-l., \& {Perrone}, M. 2010, ArXiv e-prints.
  \eprint{1004.1680}

\bibitem[{{Portegies Zwart} et~al.(2007){Portegies Zwart}, {Belleman}, \&
  {Geldof}}]{2007NewA...12..641P}
{Portegies Zwart}, S.~F., {Belleman}, R.~G., \& {Geldof}, P.~M. 2007, NewA, 12,
  641

\bibitem[{{Protopapas}(2010)}]{2010AAS...21523005P}
{Protopapas}, P. 2010, in American Astronomical Society Meeting Abstracts
  \#215, vol.~42 of Bulletin of the American Astronomical Society, \#230.05

\bibitem[{{Rodr{\'{\i}}guez-Ramos} et~al.(2006){Rodr{\'{\i}}guez-Ramos},
  {Marichal-Hern{\'a}ndez}, \& {Rosa}}]{2006SPIE.6272E..35R}
{Rodr{\'{\i}}guez-Ramos}, J.~M., {Marichal-Hern{\'a}ndez}, J.~G., \& {Rosa}, F.
  2006, in Society of Photo-Optical Instrumentation Engineers (SPIE) Conference
  Series, vol. 6272 of Society of Photo-Optical Instrumentation Engineers
  (SPIE) Conference Series

\bibitem[{{Rosa} et~al.(2004){Rosa}, {Marichal-Hernandez}, \&
  {Rodriguez-Ramos}}]{2004SPIE.5572..262R}
{Rosa}, F.~L., {Marichal-Hernandez}, J.~G., \& {Rodriguez-Ramos}, J.~M. 2004,
  in Society of Photo-Optical Instrumentation Engineers (SPIE) Conference
  Series, edited by {J.~D.~Gonglewski \& K.~Stein}, vol. 5572 of Society of
  Photo-Optical Instrumentation Engineers (SPIE) Conference Series, 262

\bibitem[{{Sainio}(2010)}]{2010CoPhC.181..906S}
{Sainio}, J. 2010, Computer Physics Communications, 181, 906

\bibitem[{{Sanders} \& {Kandrot}(2010)}]{SandersKandrot}
{Sanders}, J., \& {Kandrot}, E. 2010, {CUDA by Example: An Introduction to
  General-Purpose GPU Programming} (Addison-Wesley Professional)

\bibitem[{{Schaaf} \& {Overeem}(2004)}]{2004ExA....17..287S}
{Schaaf}, K., \& {Overeem}, R. 2004, ExA, 17, 287

\bibitem[{{Schive} et~al.(2010){Schive}, {Tsai}, \&
  {Chiueh}}]{2010ApJS..186..457S}
{Schive}, H.-Y., {Tsai}, Y.-C., \& {Chiueh}, T. 2010, \apjs, 186, 457

\bibitem[{{Spurzem, R. and others}(2010)}]{spurzem2010}
{Spurzem, R. and others} 2010, in 2010 IEEE 10th International Conference on
  Computer and Information Technology (CIT'10), 1189

\bibitem[{{Szydlarski} et~al.(2011){Szydlarski}, {Esterie}, {Falcou},
  {Grigori}, \& {Stompor}}]{2011arXiv1106.0159S}
{Szydlarski}, M., {Esterie}, P., {Falcou}, J., {Grigori}, L., \& {Stompor}, R.
  2011, ArXiv e-prints. \eprint{1106.0159}

\bibitem[{{Thompson} et~al.(2010){Thompson}, {Fluke}, {Barnes}, \&
  {Barsdell}}]{2010NewA...15...16T}
{Thompson}, A.~C., {Fluke}, C.~J., {Barnes}, D.~G., \& {Barsdell}, B.~R. 2010,
  NewA, 15, 16

\bibitem[{{Venkatasubramanian}(2003)}]{2003cs.......10002V}
{Venkatasubramanian}, S. 2003, ArXiv e-prints. \eprint{cs/0310002}

\bibitem[{{Wong} et~al.(2009){Wong}, {Wong}, {Feng}, \&
  {Tang}}]{2009arXiv0908.4362W}
{Wong}, H.-C., {Wong}, U.-H., {Feng}, X., \& {Tang}, Z. 2009, ArXiv e-prints.
  \eprint{0908.4362}

\bibitem[{{Zink}(2011)}]{2011arXiv1102.5202Z}
{Zink}, B. 2011, ArXiv e-prints. \eprint{1102.5202}

\end{thebibliography}

\end{document}